\documentclass[5p,times,preprint]{elsarticle}
\RequirePackage{lineno}

\usepackage[colorlinks=true,linkcolor=blue,citecolor=blue]{hyperref}
\biboptions{sort&compress}
\DeclareGraphicsExtensions{.pdf,.jpeg,.png}

\usepackage{url}
\usepackage{comment}
\usepackage{caption}
\usepackage{subcaption}
\usepackage{enumitem}
\usepackage{booktabs}
\usepackage{lipsum}

\begin{document}

\begin{frontmatter}
\def\papertitle{Dose Prediction with Deep Learning for Prostate Cancer Radiation Therapy: Model adaptation
to Different Treatment Planning Practices}

% please a separate Acknowledgement section at the end of the paper with content as: %We would like to thank the National Institutes of Health (NIH) for supporting this %study through a research grant (R01CA237269) and Dr. Jonathan Feinberg for editing %the manuscript..
 \title{\papertitle}

\author[1,2]{Roya Norouzi Kandalan}
%\ead{RoyaNorouzi@my.unt.edu}
\author[1]{Dan Nguyen}
%\ead{Dan.Nguyen@UTSouthwestern.edu}
\author[1]{Nima Hassan Rezaeian}
%\ead{Nima.Hassan-Rezaeian@UTSouthwestern.edu}
\author[3]{Ana M. Barragán-Montero}
%\ead{ana.barragan@uclouvain.be}
\author[4]{Sebastiaan Breedveld}
%\ead{s.breedveld@erasmusmc.nl}
\author[2]{Kamesh Namuduri}
%\ead{kamesh.namuduri@unt.edu}
\author[1]{Steve Jiang \fnref{cor1}}
\ead{Steve.Jiang@UTSouthwestern.edu}
\author[1]{Mu-Han Lin \fnref{cor1}}
\ead{Mu-Han.Lin@UTSouthwestern.edu}
%\lipsum[1-2]

\cortext[cor1]{Co-corresponding authors}

\address[1]{Medical Artificial Intelligence and Automation (MAIA) Laboratory, Department of Radiation Oncology, UT Southwestern Medical Center 2280 Inwood RD EC3.160J, Dallas, TX 75390-9303, USA}
\address[2]{Department of Electrical Engineering, University of North Texas, Denton, TX, USA}
\address[3]{Molecular Imaging, Radiotherapy and Oncology (MIRO), UCLouvain, Brussels, Belgium}
\address[4]{Department of Radiation Oncology, Erasmus University Medical Center – Cancer Institute, Rotterdam, The Netherlands}

%Please follow the green journal style for co-corresponding %authors' symbol, address, emails, etc. see this paper as an %example: https://www.thegreenjournal.com/action/showPdf?%pii=S0167-8140%2819%2933111-1 

\begin{abstract}	% <= 250 words

\textbf{Purpose}: This work aims to study the generalizability of a pre-developed deep learning (DL) dose prediction model for volumetric modulated arc therapy (VMAT) for
prostate cancer and to adapt the model, via transfer learning with minimal input data, to three different internal treatment planning styles and one external institution planning style.     

\textbf{Methods}: We built the source model with planning data from 108 patients previously treated with VMAT
for prostate cancer. For the transfer learning, we selected patient cases planned with three different styles, 14-29 cases per style, in the same institution and 20 cases treated in a different institution to adapt the source model to four target models in total. We compared the dose distributions predicted by the source model and the target models with the corresponding clinical dose predictions used for patient treatments and quantified the improvement in the prediction quality for the target models over the source model using the Dice similarity coefficients (DSC) of 10\% to 100\% isodose volumes and the dose-volume-histogram (DVH) parameters of the planning target volume and the organs-at-risk. 

\textbf{Results}: The source model accurately predicts dose distributions for plans generated in the same source style, but performs sub-optimally for the three different internal and one external target styles, with the mean DSC ranging between 0.81-0.94 and 0.82-0.91 for the internal and the external styles, respectively. With transfer learning, the target model predictions improved the mean DSC to 0.88-0.95 and 0.92-0.96 for the internal and the external styles, respectively. Target model predictions significantly improved the accuracy of the DVH parameter predictions to within 1.6\%.  

\textbf{Conclusion}: We demonstrated the problem of model generalizability for DL-based dose prediction and the feasibility of using transfer learning to solve this problem. With 14-29 cases per style, we successfully adapted the source model into several different practice styles. This indicates a realistic way forward to widespread clinical implementation of DL-based dose prediction.

\end{abstract}
% \maketitle
\end{frontmatter}

% \keywords{}

% \linenumbers\modulolinenumbers[5]

\section{Introduction}

Radiation therapy treatment planning is a complex process, as the target dose prescription and normal tissue sparing are conflicting objectives. The lowest achievable dose for each individual organ-at-risk (OAR) is unknown a priori. Multiple iterations between the planner and the physician may be required to reach the optimal balance between target coverage and OAR sparing. Recent planning tool developments have focused on improving the efficiency of the planning iteration process. Interactive multi-criteria optimization (MCO) allows the planner or the physician to explore the trade offs among the possible solutions and eliminates the communication lag in the plan iteration process \cite{Craft2012, Monz08, osti_20853459,havij, BREEDVELD20191}. Knowledge-based planning (KBP) takes a data-driven approach to learn from the past high-quality clinical plans and predicts a patient's specific dose-volume histogram (DVH) and OAR dose constraints for new patients by using the relationship between geometric and dosimetric information derived from historical patient plans in the library \cite{ Wu_2009, Wu_2011, Yuan_2012, Appenzoller_2012, Wu_2014, Moore_2011, Shiraishi_2016, Wu_2013}. 

Advancements in the field of artificial intelligence (AI), especially in data-driven machine learning (ML) and deep learning (DL) algorithms for many challenging computer vision problems, have inspired many researchers in radiation oncology. Deep learning and, more specifically, convolutional neural network (CNN) architectures have significantly improved imaging and vision tasks. In particular, UNet \cite{Ronneberger_2015}, a complex architecture initially designed for biomedical image segmentation, has notably improved the performance of predicting the radiation dose distribution in the body without going through a real planning process \cite{Nguyen_2019,Siri2019,  Nguyen2019,Montero2019}. The predicted dose distribution provides both visual input and DVH metrics to assist physician's trade-off decision-making up front to provide more achievable planning objectives prior to the planning process. 
%some issues when citing dose prediction papers. 1. Penelope's %paper is about dose calculation not dose prediction; 2. don't %cite abstracts; 3. cite both arxiv and published papers. 

While DL models holds promise as accurate predictors of the expected dose distribution prior to planning, the heterogeneity of data is a common challenge for AI modeling. Despite nationally-accepted practice guidelines that set the baseline of patient care, actual clinical practice is rarely clearly defined in black and white. Patient-specific medical reasons such as hip prosthesis and previous radiation can lead to deviation from the guideline. Strategies for solving a specific dosimetric trade-off problem will vary among different physicians and planners \cite{Chan2018}.

Treatment planning systems and optimization algorithms also introduce variations into clinical practice. Plans generated by different practice styles that meet the national guidelines, in terms of plan quality, can end up with different spatial dose distributions. A DL dose prediction model built based on a particular dataset from one institution may not work well for different treatment planning styles even within the same institution. Therefore, the ability to adapt a pre-built DL dose prediction model to a given planning/practice style is desirable. 

Furthermore, expertise in AI modeling is not widely accessible in various clinical settings in the field of radiation oncology. Although some AI solutions are commercially available and the vendor provides AI expertise, building an initial model still requires a clean and large dataset of treated patient cases. This requires a tremendous effort from the user to collect and curate enough data for modeling. Sharing models between institutions could alleviate the challenges associated with building an initial AI model, but the heterogeneity in practice mentioned above would still
yield unsatisfactory performance. Thus, the ability to easily adapt a pre-built dose prediction model to a different practice style could lead to practical clinical implementation of DL dose prediction models in real-world clinical settings.

The goals of this work are to investigate the problem of generalizing DL-based dose prediction models and to utilize transfer learning to adapt a DL prostate VMAT dose prediction model to various planning/practice styles with minimal
data from each individual style. We built a source model based on the 108 patients treated with VMAT for prostate cancer in a large institution.This work tests two types of practice heterogeneity: first, different planning styles in the same institution where we built the source model; and second, planning practices at a different institution. The source model was adapted to three internal planning styles and one external planning style with 14-29 training cases for each model. To our knowledge, this is the first work to study the DL-based dose prediction model generalizability problem and to utilize transfer learning to provide a practical solution.

\section{Methods}
\subsection{Data Groups}
For this study, we selected a total of 248  cases of prostate cancer treated with the VMAT technique. 188 cases from one institution, planned with the Eclipse treatment planning system (TPS) utilizing the progressive resolution optimizer for VMAT optimization; and the 60 cases from an external institution, planned with the Erasmus-iCycle TPS \cite{van_Haveren_2019}, a system for automated multi-criteria treatment planning. Table \ref{tbl:data} summarizes the datasets and shows the dose distributions of individual styles.

The source model was trained with the "Source" dataset, which consists of 108 plans in the ‘conformal’ dose style, which is the most representative style of planning in our institution. In the same institution, we found three additional planning styles for prostate cancer VMAT treatments, represented by three target datasets: "Internal-A," "Internal-B," and "Internal-C".  The Internal-A and Internal-B styles are slightly more aggressive in OAR sparing than the "Source" style.  The Internal-C style is a more extreme approach that allows a higher dose to femoral heads to spare the bladder and rectum. We use these three target datasets to demonstrate the problem of model generalizability and to test the feasibility of using a small dataset for transfer learning by using 14-29 cases from each target dataset. We use 2-5 test cases to evaluate the target model's performance.   

The "External" style represents plans from a different institution. As one can see, the dose distribution of the External style has many intermediate dose spikes in both the lateral and the anterior directions. Given the large differences between the External and the Source styles, we trained the target model with 20 cases and thoroughly evaluated it with 40 cases. 

\begin{table}[h!]
     \begin{center}
     \begin{tabular}{ c  p{1.5cm}  p{1cm}  p{1cm}   p{1cm} p{1cm}  }
     \toprule
      Dose Style & Name & Dataset & Training & Testing  \\ 
    \cmidrule(r){1-1}\cmidrule(lr){2-2}\cmidrule(l){3-3}\cmidrule(l){4-4}\cmidrule(l){5-5}\cmidrule(l){6-6}
     \raisebox{0cm}{\includegraphics[width=0.1\textwidth]{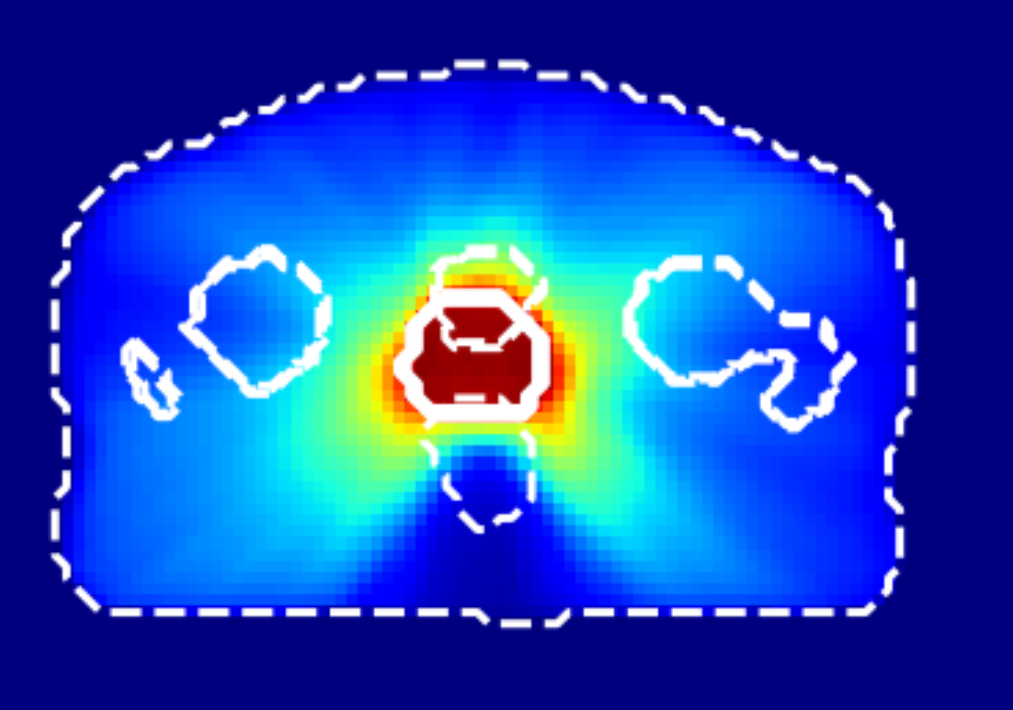}} & Source & 118& 108 & 10 
      \\
      \raisebox{0cm}{\includegraphics[width=0.1\textwidth]{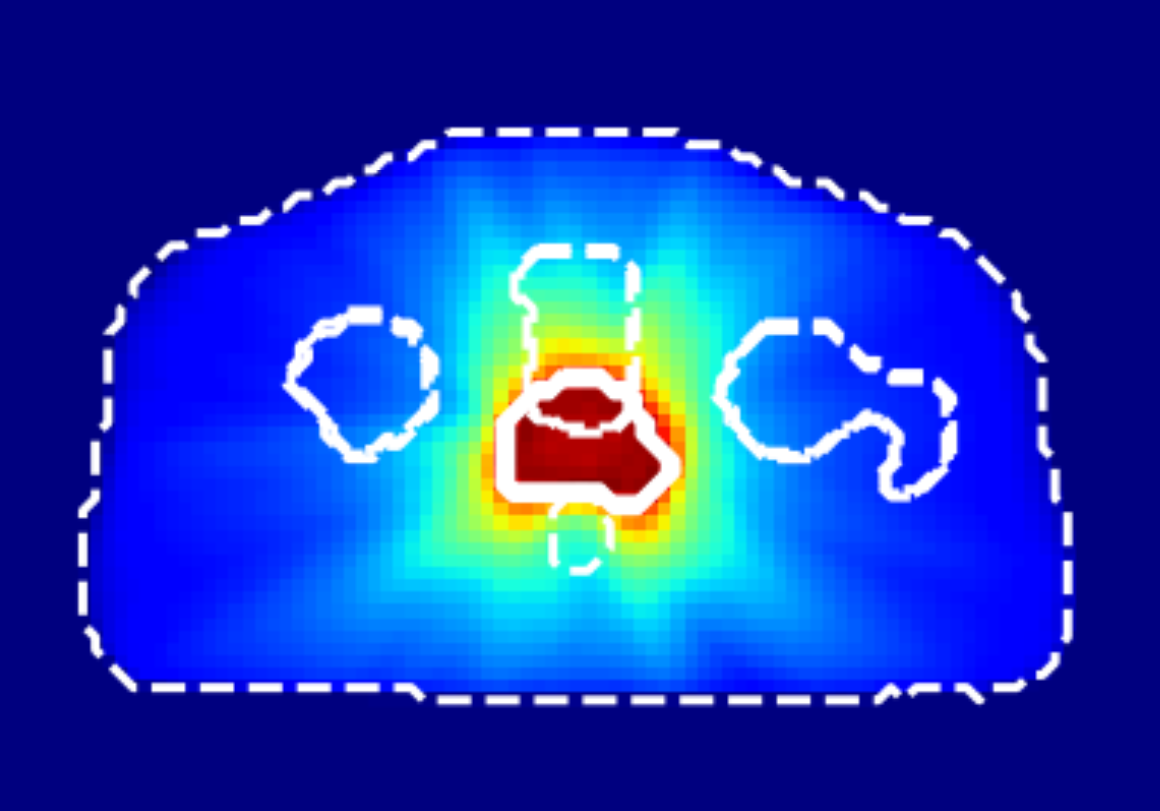}} & Internal-A & 34& 29 & 5
      \\
      \raisebox{0cm}{\includegraphics[width=0.1\textwidth]{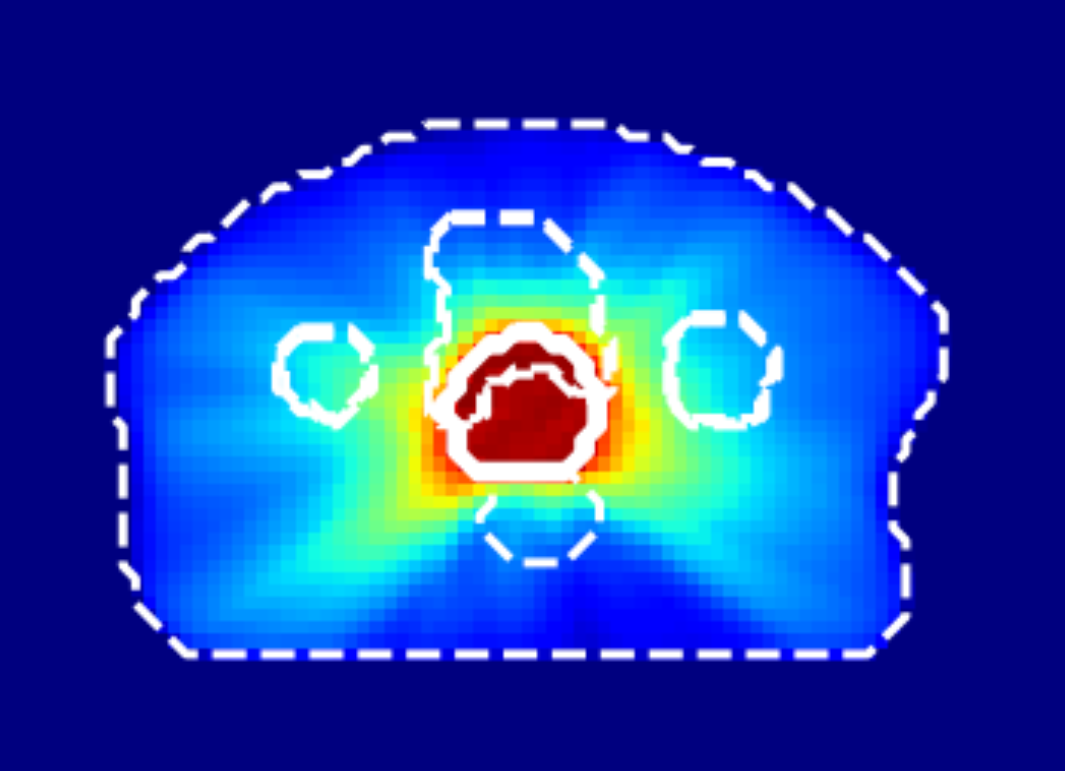}} & Internal-B & 16  & 14 & 2
      \\
      \raisebox{0cm}{\includegraphics[width=0.1\textwidth]{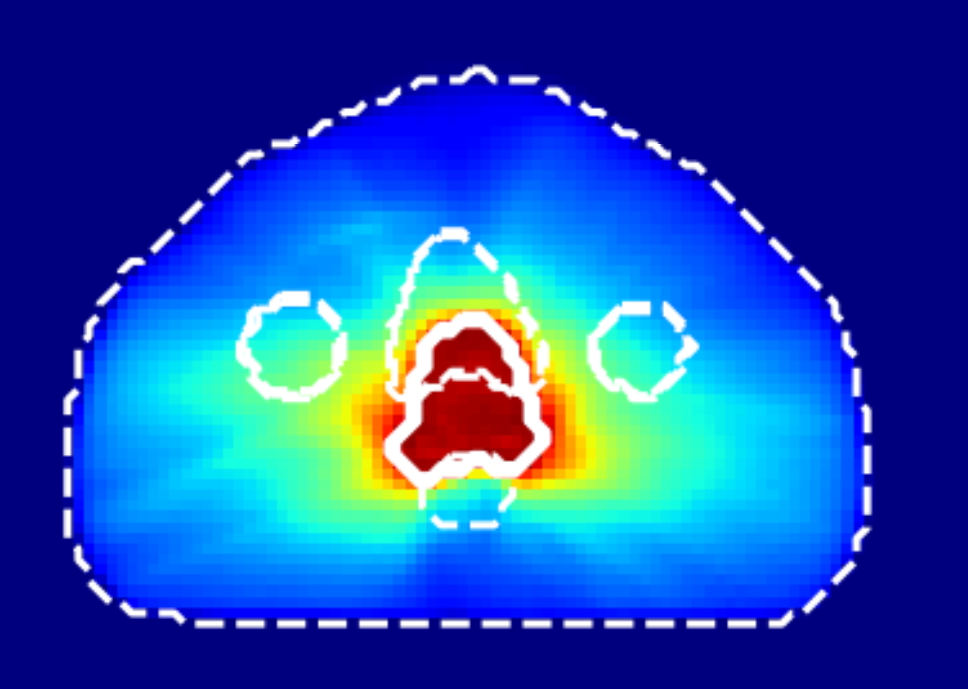}} & Internal-C & 20 & 17 & 3
      \\
      
      \hline
      \raisebox{0cm}{\includegraphics[width=0.1\textwidth]{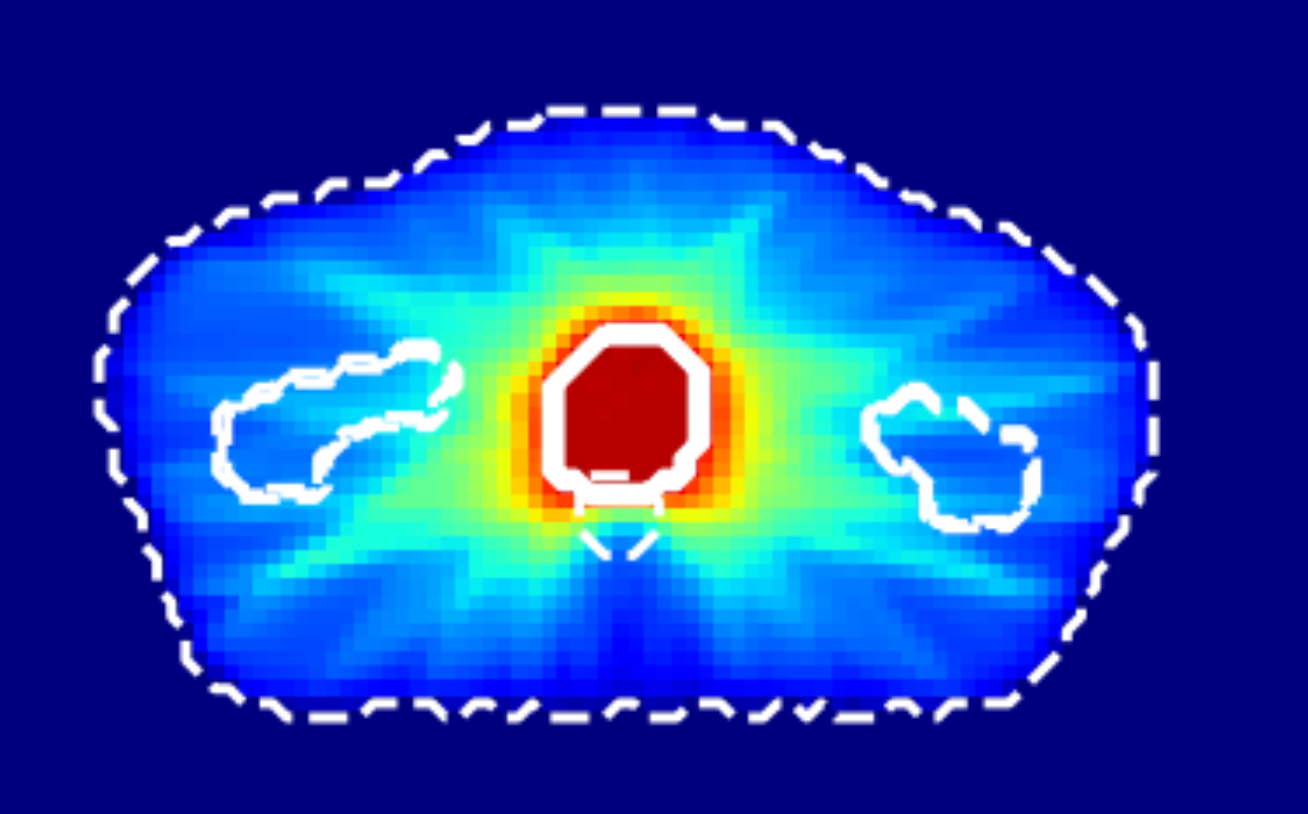}} & External & 60  & 20 & 40 
      \\
      \bottomrule
      \end{tabular}
      \caption{Patient datasets used for building the source model (Source) and for testing the model generalizability and then for adapting model (Internal-A, Internal-B, Internal-C, and External). First column shows a typical dose distribution of the planning style represented by each dataset.}

      \label{tbl:data}
      \end{center}
      \end{table}

\subsection{Network}
In this work, we implemented a 3D UNet for the model architecture, \cite{Ronneberger_2015}, as UNet has been extensively used in radiation therapy for dose prediction \cite{Nguyen_2019,Siri2019, Nguyen2019,Montero2019}. The inputs to this architecture are the contours of the planning target volume (PTV) and OARs, including bladder, rectum, left and right femoral heads, and body, which were formatted as binary masks with a size of $[120 \times 120 \times64]$ each. Due to the memory limits of a 16 GB GPU (NVIDIA K80), we implemented patch-based training, with a patch size of $[96 \times 96 \times 48]$. At each training iteration, the patches were randomly selected from the binary mask data based on a Gaussian sampling scheme proposed by \cite{Nguyen17}. We also implemented group normalization \cite{wu2018group} after each set of convolution and rectified linear unit (ReLU) operations, which helps accelerate the convergence rate in the network. This model is trained to learn the mapping between the binary masks of the OARs and the clinical radiation dose distribution in the body. Mean Square Error (MSE) was the loss function, and the Adam optimizer with a learning rate of $10 ^{-4}$ was used to minimize the MSE. The final network consists of 85 layers with 7,870,177 trainable parameters.  

\subsection{Transfer learning}
For both the internal Source dataset and the external institution dataset, the transfer learning method implemented for style transfer involves 3 components: 1) freezing the weights of the first half of the model, 2) training only the second half of the model, and 3) randomly reinitializing the weights of the very last layer of the model. Assuming that the major variations in dose distributions are owing to different planning styles across institutions, we can effectively freeze the first half of the UNet, which finds features from just the anatomy. We can then train just the second half of the network to learn the dose distributions of a different planning style. Finally, to assist the network in converging and to prevent it from falling into a local minimum, we reinitialize the weights in the final layer of the network. The hyperparameters remain unchanged, except for the learning rate, which was reduced to one-tenth of the original value.

\subsection{Model Evaluation}
We first applied the Source model to the Source test cases to evaluate the model's performance on its own planning style. We then applied the Source model to the test cases of the three internal and one external target styles to investigate the model's performance on other planning styles and to set the baseline prediction quality before transfer learning. Individual target models trained via transfer learning were each applied to the test cases of their own planning style to investigate the prediction quality after adapting the Source model to the given target style. To further evaluate the improvement achieved by the transfer learning, we cross-compared each of the doses predicted by the Source and target models with the respective clinical plan dose.   

We compared the absolute doses predicted by the DL models to the clinical plan doses without further normalization. We evaluated the spatial dose distributions and the dose-volume histograms (DVH). We calculated DVH metrics including structure mean and maximum doses (i.e. D2), PTV D98, and PTV D95. The clinical plan dose served as the baseline for the dose comparison. We calculated the differences in the DVH metrics between the clinical plan dose and the predicted doses to quantify the DVH agreement. In addition, we calculated the dice similarity coefficients (DSC),  $\frac{2(A \cap B)}{(A+B)}$, for isodose volumes from 10\% to 100\% of the prescribed dose to quantify the agreement of the spatial dose distribution between the clinical plan dose and the predicted doses. We used a paired t-test to calculate the statistical significance of the results for the External style, which had enough test cases for a meaningful test.

\section{Results}

\begin{figure*}
    \centering
    \includegraphics[width=\textwidth]{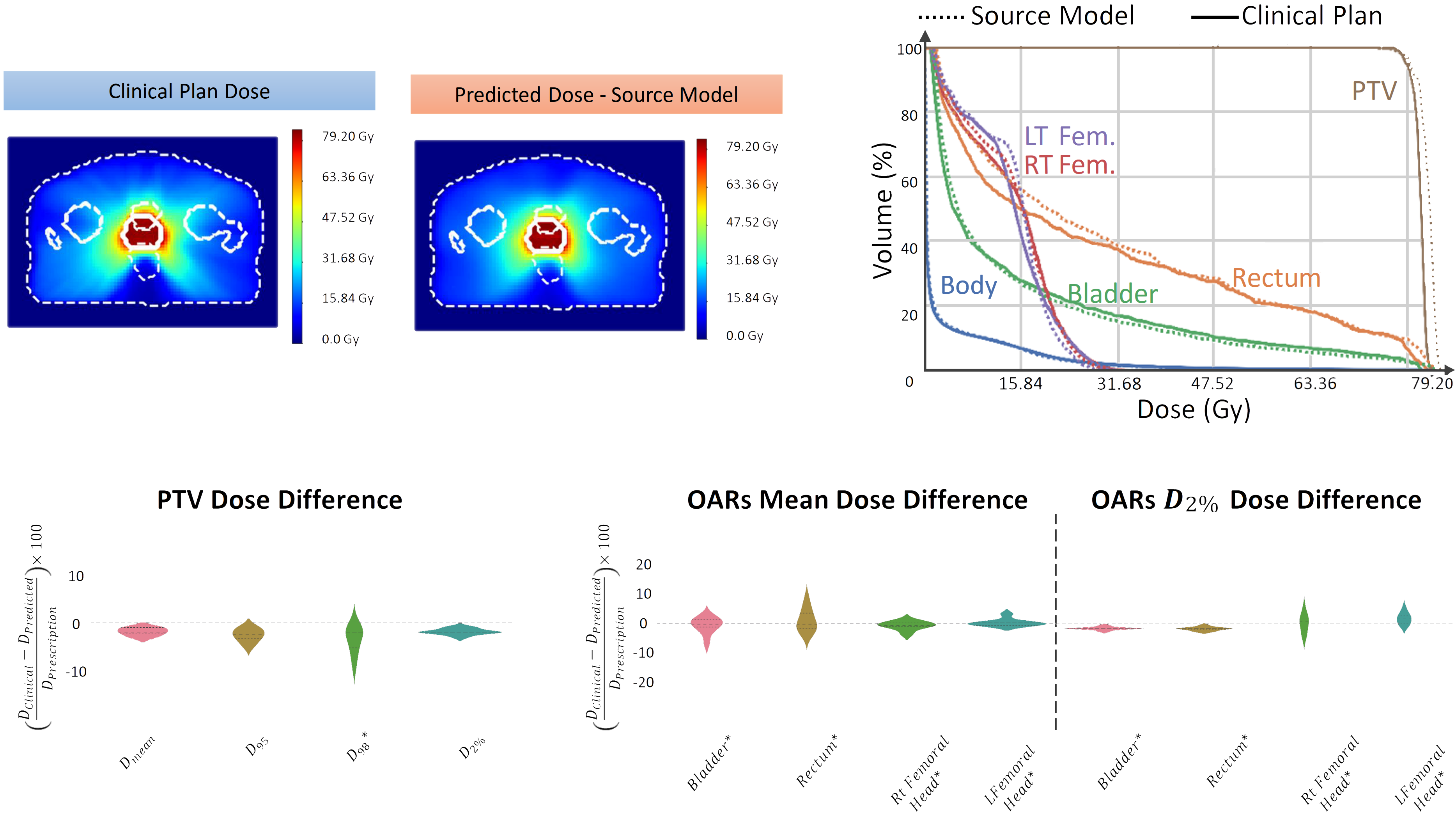}
    \caption{Performance analysis of the source model including the dose distributions with the structure contours outlined in white dotted lines; the DVH of the clinical plan (solid lines) and the predicted dose (dotted lines)l and the violin plots of the target and the OAR DVH metrics.  }
    \label{fig:source}
\end{figure*}

Figure \ref{fig:source} shows a typical example of the dose agreement between the clinical plan and the Source model prediction for the Source style test cases. Upon visually inspecting the dose distribution and the DVH, one can see that the Source model predicts the PTV dose correctly with minimal differences in the OAR doses. The violin plots present the differences in the PTV and the OAR DVH metrics between the clinical plan and the Source model prediction. The differences in the PTV DVH metrics are mostly within 3\%, as reflected by the width of the individual data clusters. The median of the PTV dose differences are all within 2\%. Similarly, the median of the mean and maximum OAR dose differences are within 1\% and 2\%, respectively. 

\begin{figure*}
    \centering
    \includegraphics[width=\textwidth]{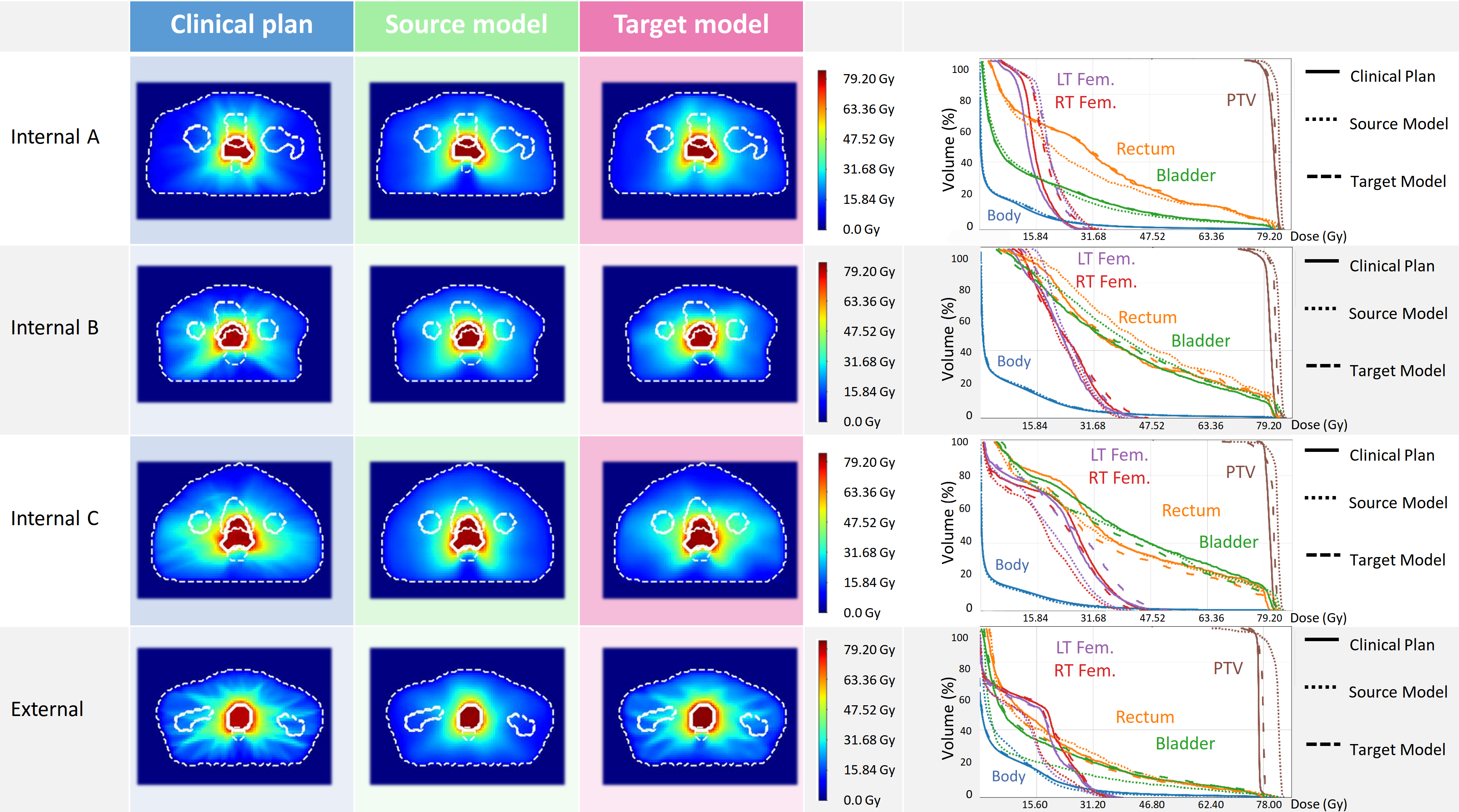}
    \caption{Dose comparison between the clinical plan and the predicted plan using the source and the target models including the dose distributions and the DVH comparisons.}
    \label{fig:dvh}
\end{figure*}

Figure \ref{fig:dvh} illustrates the Source model and target model predictions on the test cases of individual target styles. The improvement in the model performance after transfer learning can be seen in the dose comparisons among the clinical plan, the Source model prediction and the target model prediction. The Source model fails to predict the style-specific dose distribution features. However, the target models have learned those dose distribution features via transfer learning, so they predict a distribution similar to the clinical plan and exhibit a better agreement in the DVH, especially for the Internal-C and External styles. 

\begin{figure*}
    \centering
    \includegraphics[height=10cm]{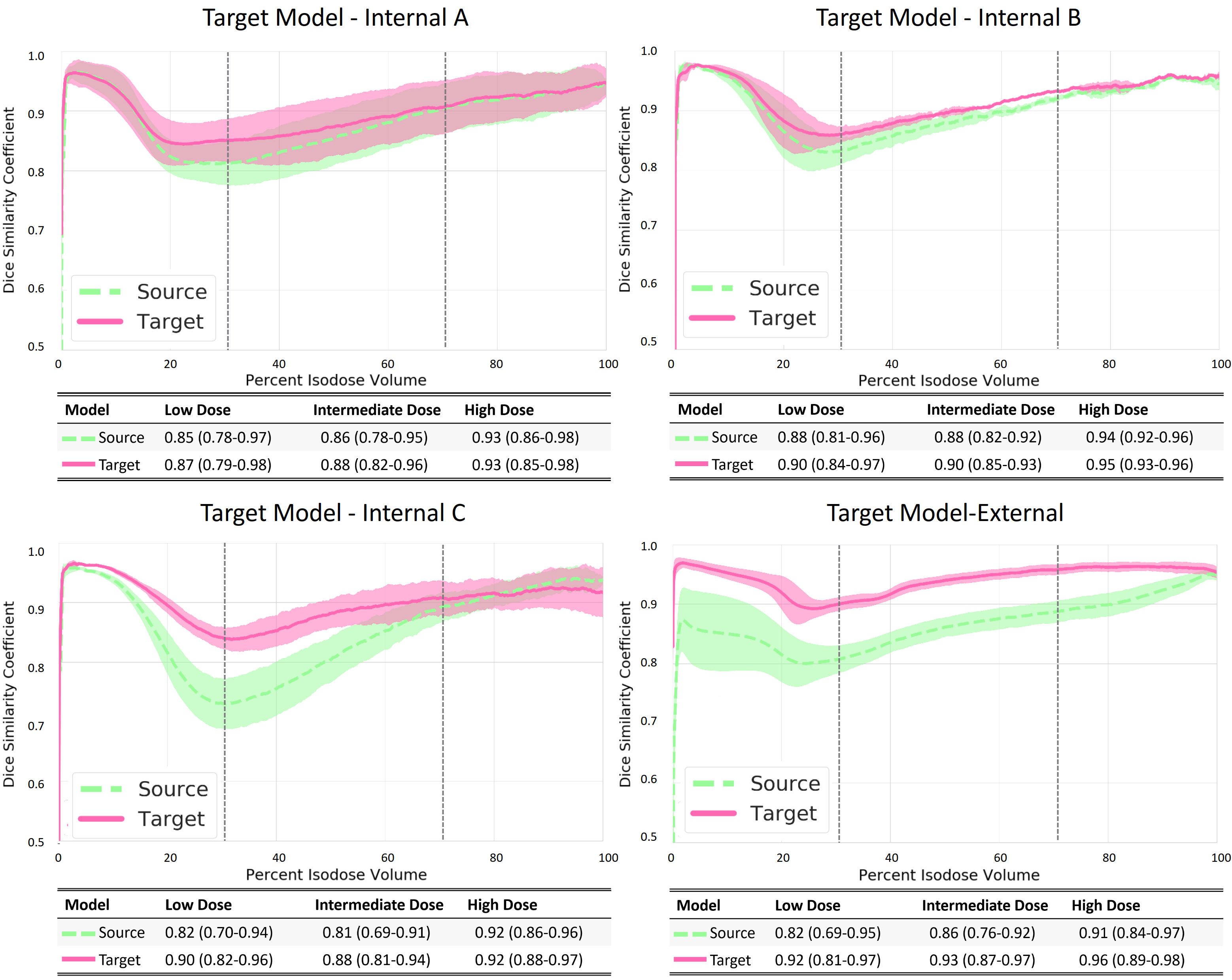}
    \caption{Dice similarity coefficients analysis, \big($\frac{2 (A\cap B)}{A+B}$\big).
    Comparing the isodose volumes between the clinically planned dose and the predicted dose. One standard deviation represented the error in the graphs.  }
    \label{fig:dsc}
\end{figure*}
We compared each of the Source and target model predictions to the respective clinical plan dose and calculated the DSC of individual isodose volumes accordingly. The DSC values increased with the dose agreement with a coefficient of 1 considered to be a perfect match. As presented in Figure \ref{fig:dsc}, the target models improved upon the Source model in term of DSC regardless of the planning style. However, the Internal and the External styles exhibit different trends. For the Internal styles, both the Source and the target models predict the high dose bath (70\% to 100\%) fairly well. The main improvements appear in the intermediate dose bath (60\% to 30\%). The mean DSC ranges between 0.81-0.94 for the three Source models and between 0.82-0.91 for the corresponding target models. The largest improvement may be seen in the Internal-C style, as the target model achieves a 7\% improvement over the Source model (mean DSC is 0.80 for the Source model and 0.87 for the Internal-C target model) transfer learning. For the External style, the target model improves significantly upon the Source model throughout all isodose volumes ($p<0.05$), with the improvement increasing as the isodose volume decreases. An average of 5\% and 8\% mean DSC improvements were achieved in the high and the intermediate dose volumes, respectively. A systematic 10\% improvement was seen in the low dose volume. 

\begin{figure*}
    \centering
    \includegraphics[height=10cm]{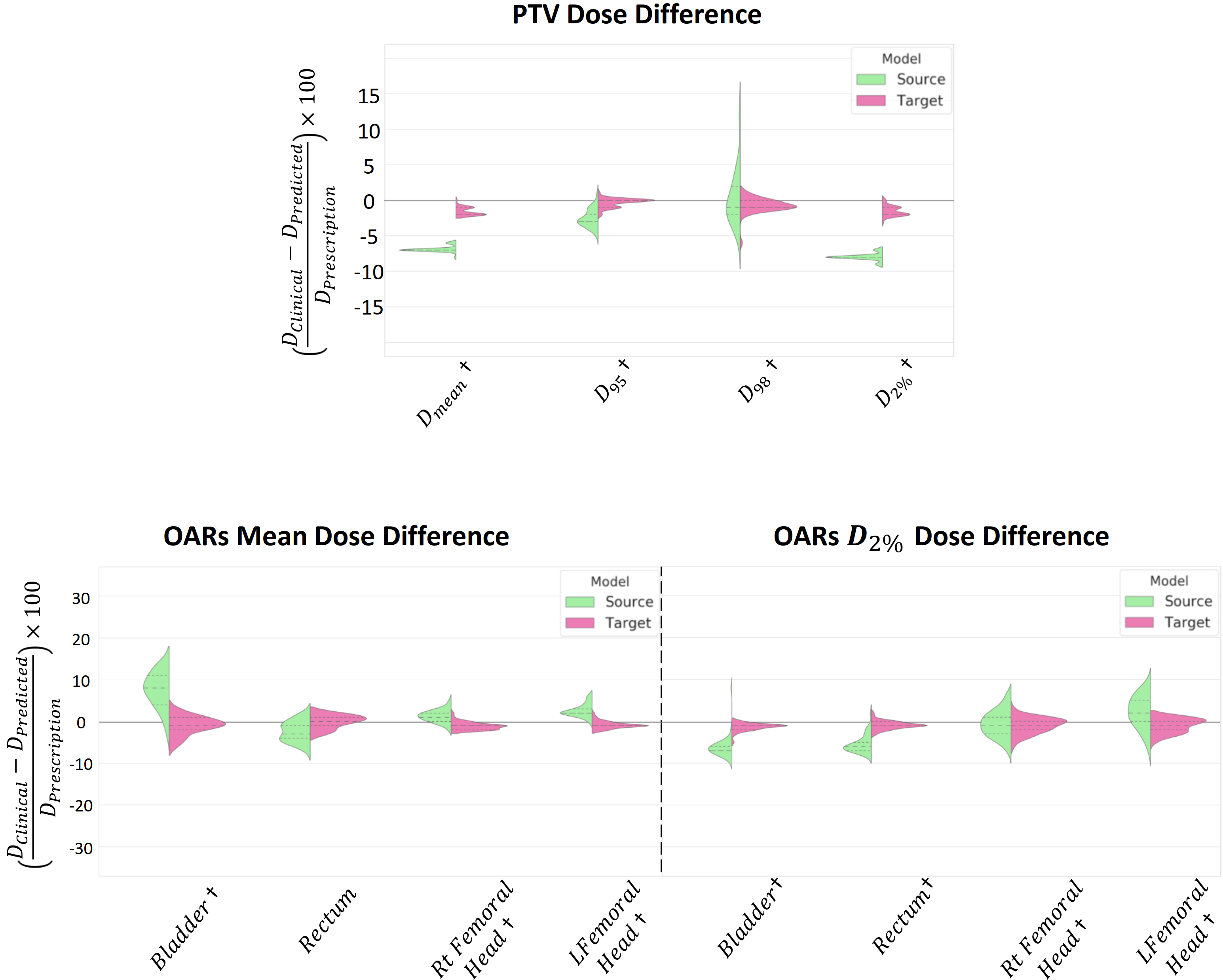}
    \caption{Quantitative performance evaluation of the error in the DVH metric in target model dose prediction vs the source model dose prediction in respect to the clinically planned dose for the external data.\\
    $ ^{\dagger}$ Represents $p< 0.05$\\
    }
    \label{fig:violin}
\end{figure*}

Figure \ref{fig:violin} illustrates the Source and the External target model performances on predicting the DVH metric of the External style test cases. Overall, as indicated by the length of the data clusters, the Source model demonstrates a large variation in the prediction quality, while the External target model demonstrates more consistent and accurate predictions. The Source model overestimates the PTV doses, especially for the Dmean and the D2. The External target model significantly improved the predictions and resulted in mean differences within 1.6\% (Table \ref{tbl:E}). In terms of the OAR dose predictions, the target model achieved an agreement within 1.5\%. In summary, the External target model improved upon the Source model by up to 6.4\% and 6.0\% in the PTV and the OAR DVH predictions, respectively ($p<0.05$).

\begin{table}
  \caption{Quantitative performance evaluation based on the error in target model dose prediction vs the source model dose prediction. Error is calculated in respect to the clinically planned dose for the external data.}
  \label{tbl:E}
  \includegraphics[width=\linewidth, scale=0.5]{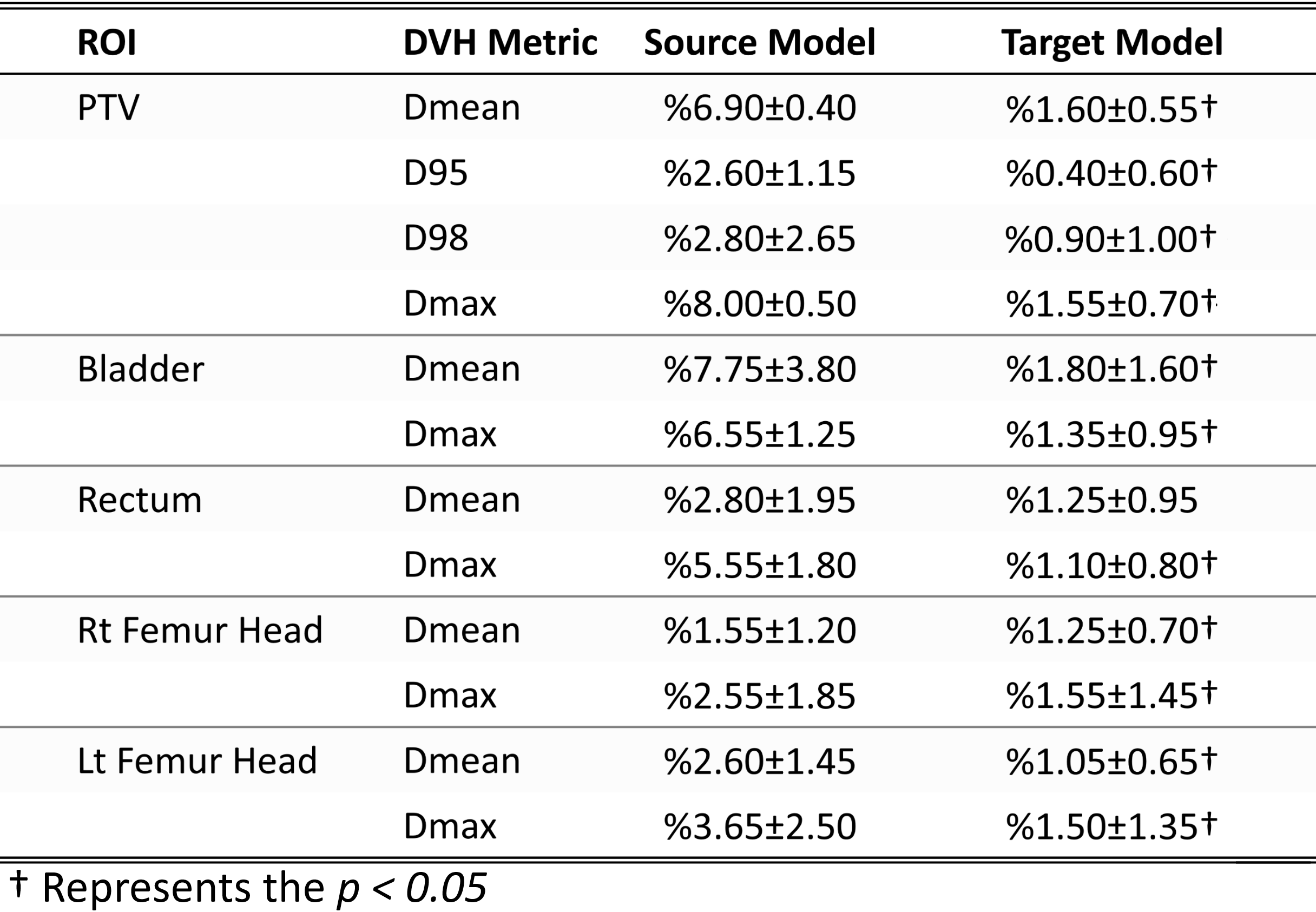}
\end{table}

\section{Discussion and Conclusion}
\begin{comment}
echo the introduction
data heterogeneity problem, planing styles, external data, small clinics
\end{comment}

There has been growing attention on leveraging AI-based decision support tools (DST) to improve the treatment planning process. Incorporating AI DST as a part of the clinical pathway 
\cite{Craft2012} could standardize evidence-based practices to ensure high quality and cost-effective medical care. AI algorithms, including the DL dose prediction model in this work, require large repositories of high quality data. This becomes one of the barriers to widespread deployment of AI-based solutions in the radiation oncology field. In this study, we utilized transfer learning to solve the data size problem and demonstrated the ability to adapt a source model to three different internal planning styles and one external planning style with minimal data input. 

Heterogeneity in clinical practices, whether intra- or inter-institutional, is commonly seen. The clinical protocols and guidelines set the floor of treatment plan acceptance, but users set the ceiling for their own practice. Table \ref{tbl:data} clearly shows that, even within the same institution, the individual trade-off preferences of different physicians and planners in combination result in a variety of dose distribution styles. On top of these differences in practice, the different treatment planning system employed in the External institution in this study yielded dramatically different dose distributions from the Source style. The treatment plans in the Source style were optimized with a specific objective function (normal tissue objective, NTO) to shape the dose fall-off conformally, whereas plans from the External institution were optimized with a unique 23-beam starting condition of the VMAT optimization that led to distinct spikes in the low dose bath. Regardless of the dose distribution style, they are all clinically accepted plans that comply with the clinical guideline. The AI model’s adaptation to the user’s practice style can achieve standardization with respect to the user's prior experience. This would be practical and beneficial for implementing AI-guided planning in practice, because the user can precisely interpret the predicted dose for meaningful clinical use. 

Directly deploying a model built in a different practice style may lead to unsatisfactory predictions. As shown in the Source model predictions on the cases planned with target styles in Figure \ref{fig:dvh}, the predicted doses inherit the conformal dose distribution of the Source style. They meet the planning objectives, but they fail to represent the dose distribution features of the target styles. This reflects a common frustration of institutions that are trying to clinically implement a model provided by a vendor or developed by a different institution. In this work, we demonstrated the transfer learning can adapt a source model into various practice styles.    

Transfer learning with an additional 14-29 cases in the target style, allowed the target models to learn the features of the new planning style quickly. Taking the External style as an example, the distinct low dose spikes were precisely predicted by the External target model. We saw this improvement globally in all target styles, and it increased as the variation between the Source and target styles increased (Figure \ref{fig:dvh}). For example, the Internal-A and Internal-B styles pull the dose from the rectum only slightly more aggressively than the Source style, so the inherent differences are small. Therefore, the Source model still achieved more than a 0.78 mean DSC, which the target models only slightly outperformed. In contrast, the Internal-C style turns off the NTO and intentionally trades femur doses for lower rectum and bladder doses, so it differs
more substantially from the Source style. Accordingly, the Source model predictions on the Internal-C style cases demonstrated a significant dip in DSC in the intermediate dose levels, but the Internal-C target model improved the DSC to a satisfactory level. For the External style, which is fundamentally different from the Source style, the External target model improved upon the Source model’s DSC by up to 10\% DSC and had a DSC higher than 0.87 among all test cases and isodose volumes. 

The intermediate dose bath of the VMAT plan is more unpredictable because of its complex nature. The variations in trade-offs and planning approaches can tremendously polarize the spatial dose distribution in this dose range. Moreover, with the large numbers of beam angles in VMAT optimization, the optimizer has more options to pull the dose off a specific area, which makes the dose more uncertain to predict. Traditionally, the DL dose predictor has a hard time handling these style-specific features. This is reflected in figure 
\ref{fig:dsc}, which shows that the DSC of the Source model prediction ranges between 0.80-0.94. However, with transfer learning, the target models can represent these dose features, and the DSC of target model prediction increased to 0.87-0.99, which is satisfactory.
 
In light of the barriers to accessing the expertise of AI modeling, collecting sufficient data to train a model, and sharing data among institutions, a broad AI tool that can easily adapt a maturely built model to different practice styles could allow more practices to access and clinically implement an AI model for dose prediction. We used prostate VMAT as an example to demonstrate the feasibility of leveraging transfer learning for model adaptation. However, this methodology can be employed for other AI tools in the field of radiation oncology and allow for widespread application of such tools.

\section*{Acknowledgement}

We would like to thank the National Institutes of Health (NIH) for supporting this study through a research grant (R01CA-237269) and Dr. Jonathan Feinberg for editing the manuscript.

 %\newcommand{\newblock}{} %required for natbib iopart compatibility
% \bibliographystyle{dcu}
%\newpage

\bibliographystyle{unsrt}
\bibliography{bib}

\textcopyright{} 
\end{document}